# Research on Innovation in China and Latin America: bibliometric insights in business, management, accounting, and decision sciences

Julián David Cortés-Sánchez [a,b]*

[a]*School of Management and Business, Universidad del Rosario, Bogotá, Colombia*

[b]*Fudan Development Institute, Fudan University, Shanghai, China*

*Corresponding author email: julian.cortess@urosario.edu.co

- Julián David Cortés-Sánchez is a principal professor at Universidad del Rosario's School of Management and Business (Colombia), and visiting scholar (2019) and invited researcher at Fudan University's Development Institute (China).

# Research on Innovation in China and Latin America: bibliometric insights in business, management, accounting, and decision sciences

This study aims to comprehend the structure of RIBM (research on innovation in business and management) in China and LAC (Latin America and the Caribbean) via co-word and institutional co-authorship networks using Scopus' bibliographic data (1998- 2018). Multiple Correspondence Analysis and Social Network Analysis were applied. Public institutions are interconnected and generate most of the advances in RIBM. RIBM boards regional and national STi policies permeated by sustainability-related factors. China is focused on IT and knowledge management for supply chain and engineering, while LAC focuses on institutional perspectives for economic development.

## 1 Introduction

The commercial and knowledge exchange between China and LAC (Latin America and the Caribbean) is increasing. Trading between both regions grew from USD$10 to USD$130 billion from 2000 to 2009 (Koleski, 2011). Leading universities from China and LAC (e.g., Universidad de Chile, Universidad de Los Andes, Instituto Tecnológico de Monterrey) are joining resources to foster communication and cooperation exchange between one another (FLAUC - Fudan Latin America University Consortium, n.d.).

Research on Innovation in Business and Management (RIBM) fulfills a critical role among that commercial and knowledge exchange. According to Fagerberg (2013) innovation studies deal with how innovation is generated and applied and also explain what its economic and social impacts are. RIBM is based mostly on one out of three clusters found on innovation studies: *organizing innovation* (Fagerberg et al., 2012). That cluster is composed by studies on management and organization, technology, business, firms, and sector/industry. Organizations that could extract benefits from RIBM, improve their capacity to adapt in a technologically challenging environment (Mueller et al., 2013; Pfeffer & Sutton, 2006; Rosenbusch et al., 2011; Zhao, 2015).

A conservative estimate points out that researchers in China have published 51,000+ articles in the field of business, management and accounting, while LAC published 24,000+ articles (Scopus, 2018). When research entities such as articles reach the thousands, bibliometric methods such as keyword co-occurrence or co-authorship networks enable researchers, entrepreneurs, government employees, and independent organizations to disentangle the underlying structure of mountains of digitalized literature (Zupic & Čater, 2015). Questions such as the following can be answered through bibliometric lenses: which are the research fronts in IRBM? Which are the knowledge bridges in RIBM? Which are the key institutions in terms of collaboration in RIBM?

Previous RIBM implementing bibliometrics and comparing different geographical regions have measured the differences on nanoscience output and impact between China and the USA (Kostoff, 2012), the market-orientation vocation of pharmaceutical research in Europe and the USA (Tijssen, 2009), and the social network and expertise of biotechnology scientists between UK and Germany (Casper & Murray, 2005), among other topics. Reframing the bibliometric studies on RIBM published by researchers from China and LAC, they have shed light on the proportion of STEM articles cited in patents (Guan & He, 2007), the inclusion of technology foresight and road-mapping in national and regional STi policy (Li, Chen, & Kou, 2017), the innovativeness measures (De Carvalho, Cruz, De Carvalho, Duclós, & De Fátima Stankowitz, 2017), industry relationships (Manjarrez, Pico, & Díaz, 2016) and business models (Ceretta, Dos Reis, & Da Rocha, 2016), among other topics. Details will be presented in section 2 "Research background."

Despite some advancements, several issues remain unattended. First, there is no comparative study on RIBM published by researchers affiliated with Chinese and LAC institutions, two regions with vibrant relevance among developing countries (Ortiz-Ospina, Beltekian, & Roser, 2018). Second, there is no broader perspective on the underlying structure of RIBM in terms of research topics and their knowledge intermediaries. The same happens for institutional collaborations between and within China and LAC. Third, the bibliographic data comes from a single source: WoS (Web of Science). Our focus differs from output and impact factors since they have been analyzed in-depth in both country and regional (i.e., LAC) spheres (Cortés-Sánchez, 2019; Merigó et al., 2016). Therefore, the two research questions that guide this study are:

- What is the underlying RIBM topic structure in China and LAC?
- What is the underlying institutional collaboration structure of RIBM in China and LAC?

To diversify the bibliographic data source with a reliable alternative to WoS, we used Elsevier's Scopus. In essence, this study aims to comprehend the structure of RIBM published by researchers in China and LAC via co-word and institutional co-authorship networks using Scopus' bibliographic data for a time frame of 20 years (1998-2018). We applied MCA (multiple correspondence analysis) and SNA (social network analysis). MCA and SNA were applied to article keywords. SNA was applied to institutional co-authorships. By achieving the objective mentioned above, this study would help private, public or independent institutions to allocate intellectual, financial, or physical capital to enhance their efforts within the RIBM agenda towards intermediate-strategic research topics, or to select and invest in further inter-institutional alliances, either formal or informal.

The rest of this paper is organized as follows: Section 2 and 3 present and discuss the research background and the methodology, respectively. Section 4 presents the results, which are discussed in Section 5, and Section 6 presents conclusions, limitations, and further research.

## 2  Research background

The research background consisted of three literature streams: (i) bibliometric insights on RIBM comparing different continents, sub-continents, or countries; and streams (ii) and (iii) discusses bibliometric insights on RIBM published by researchers affiliated with institutions from China and LAC, respectively. The reason behind focusing on researchers affiliated with institutions from China and LAC and their production on RIBM, is to explore the current knowledge-generation and collaboration capabilities for each region.

Regarding the first stream, studies have identified a consistent increase in China's research output on nanoscience and technology, which has doubled every 2.1 years, surpassing that of the US in 2012 (Kostoff, 2012) but maintaining a low level of citations (Guan & Ma, 2007). Tijssen (2009) argued that research from European pharmaceutical companies was oriented toward the US, rather than their home markets. When analyzing two biotechnology clusters in the UK and Germany, Casper and Murray (2005) found that the scientist's networks from the UK exhibited a balanced mix of industry and academic experience, while German scientists showed an academic-focused background.

An analysis of 21 disciplines in 34 countries and their revealed comparative advantage conducted by Harzing and Giroud (2014) proposed clustering into five groups: G1 (most populous English-speaking countries, The Netherlands and Israel): social sciences; G2 and G5 (France, Italy, Poland, Hungary, India, Russia, and Ukraine): physical sciences; G3 (Germany, Austria, and Switzerland): balanced research profile; and G4 (Asian Tigers and China): engineering. Subsequent studies suggested that the Global South should specialize in specific scientific fields to increase research impact (Confraria, Mira Godinho, & Wang, 2017; de Paulo, Carvalho, Costa, Lopes, & Galina, 2017).

Studies have identified the intellectual pillars of related fields to RIBM such as technology and innovation management (e.g., dynamic organizations, the innovation process, and knowledge management) (Pilkington & Teichert, 2006) and specific research interests depending on the regions, such as *organization*, *technology strategy*, *new product development*, *design and innovation*, *technology policy*, and *technological acquisition* in developed countries, and *technology policy*, *organization*, *technological acquisition*, *R&D management*, *technological change*, and *technological development* in developing countries (Cetindamar, Wasti, Ansal, & Beyhan, 2009). Choi et al. (2012), pursuing a similar aim, argued that among European countries, the UK has a comparative advantage in *social change*, Spain in *intellectual property*, the Netherlands in *technology policy*, Germany in *entrepreneurship*, and Italy in *technology transfer and commercialization*.

Regarding the second stream, when studying the relationship between science and technology in China, Guan and He (2007) pointed out that patents from sectors such as biotech, pharmaceutics, and organic chemistry cite more scientific papers. In contrast, patents from sectors such as information and communication technologies, semiconductors, and optics are more likely to cite other patents. Subsequent analyses focused on the inclusion of technology foresight and road-mapping in national and regional policy design applications in science, technology, and innovation (Li, Chen, & Kou, 2017), and recommendations for supporting emerging technologies (e.g., solar cells) (Huang, Zhang, Guo, Zhu, & Porter, 2014; Li, Zhou, Xue, & Huang, 2015).

Among the third stream, studies have followed a diversified agenda, including Schumpeterian innovation and cooperation (Lazzarotti, Dalfovo, & Hoffmann, 2011; Lopes & De Carvalho, 2012), innovativeness measures (De Carvalho, Cruz, De Carvalho, Duclós, & De Fátima Stankowitz, 2017), industry relationships (Manjarrez, Pico, & Díaz, 2016), business models (Ceretta, Dos Reis, & Da Rocha, 2016), financing of innovation (Padilla-Ospina, Medina-Vásquez, & Rivera-Godoy, 2018), social innovation (Silveira & Zilber, 2017), supply chain management (Tanco, Escuder, Heckmann, Jurburg, & Velazquez, 2018), and citation, publication, and academic–corporate collaboration (Cortés-Sánchez, 2019).

Highlighted findings of the third stream stated worldwide maturity in industrial relations to innovation system players (e.g., academic, scientific, or technological), the Brazilian case highlighted in LAC; five salient topics in financing innovation (i.e., financial constraints, funding sources, capital structure, venture capital, and financing of technology companies); and the research output in LAC concerning supply chain management has been insubstantial from a global perspective in terms of output, citations, publication in top-tier journals, and both international and corporate collaboration. Besides, it was revealed that these were based on the extensive use of journals with predatory features. Therefore, citations/documents ratio in the region is the lowest of the last decade.

In sum, studies reviewed have analyzed knowledge-intensive industries, the Global North-Global South relationship dynamics, technology and innovation management, innovativeness measures, and social and open innovation. Here, we will articulate the RIBM produced by researchers and institutions from China and LAC in an integrated framework, which goes beyond the Global-North (i.e., developed countries) emphasis of the first stream reviewed. The structure of RIBM will identify clustering similarities and particular properties regarding conceptual and institutional-collaboration factors applying MCA and SNA. Also, the bibliographic data source used (i.e., Scopus) differs from the ubiquitous WoS.

# 3 Methodology

## 3.1 Data source

We collected the bibliographic data using Scopus. Scopus contains more than 75 million documents published by more than 16 million authors (Scopus, 2019). We chose Scopus over WoS since most of the studies reviewed used WoS, thereby adding diversity to the previous findings. Also, Scopus has broader journal coverage and higher social sciences coverage (Gavel & Iselid, 2008; Mongeon & Paul-Hus, 2016).

The search query was limited to articles with the keyword *innovation* (*innovación* in Spanish or *Inovação* in Portuguese) in their title published within a time frame of 20 years (1998 to 2018) by at least one author affiliated with an institution located in China or LAC in journals within the fields of *business, management and accounting*, and *decision sciences*. The field of *decision sciences* was included as it integrates research on *information systems and management,* decisive platforms for digital innovation (Du & Mao, 2018; Liu et al., 2018; Ryu & Kim, 2018).

Table 1 summarizes the search query and the bibliometric data regarding the number of articles and journals, average citation per document, annual percentage growth, most relevant journals, authors' collaboration, and the top-ten of the most productive countries. The journal *Espacios* was excluded due to quality concerns (Cortés-Sánchez, 2019). After removing those articles, the final sample was composed of 1728 articles (China: 993; LAC: 735).

## 3.2 Methods and software

We used two non-redundant inputs for topic and collaboration structure analyses: articles' keywords and institutional co-authorships. Keywords and institutional co-authorships are two heterogeneous variables in bibliometric studies since keywords are word-information based (i.e., artificial connections), and co-authorships are collaboration based (i.e., 'real' connections) (Yan & Ding, 2012). Moreover, Yan and Ding (2012) argued that there are considerable similitudes among information networks based on citations: bibliographic coupling, citation, and co-citation networks. Taking into account the relatively low overall citations in developing regions compared to the Global-North (Cortés-Sánchez, 2019) and that co-word networks refer to similarities in lexical semantics instead of citations, it seems a comprehensive approach for examining RIBM in China and LAC.

For keywords, we conducted a co-word analysis using MCA and SNA. We also implemented SNA to the institutional co-authorships. The co-word analysis draws the conceptual structure of a given subject based on the co-occurrence of keywords in a bibliographic collection (Callon et al., 1983). For instance, if the keywords *innovation, patents,* and *integrated circuits*, were identified in a given document, the three are linked due to their co-occurrence in that document (Callon et al., 1983). Instead of using only author's keywords, we used the association of such keywords with *KeyWords Plus*. An automated generated key-terms based on the articles title and the references cited, thus avoiding blurry keywords such as: *article, spss, regression analysis,* and others (Clarivate Analytics, 2018). Key-terms have to appear more than twice to be considered, thus assuring the importance of the terms computed via their title-references similarity.

The MCA processes and identifies keywords that express common concepts by performing a homogeneity analysis for obtaining a low-dimensional Euclidean representation. The resulting keywords are plotted in a two-dimensional map, which shows their relative position and distribution along the dimensions, therefore the closer the keywords, the more similar in distribution (Cuccurullo et al., 2016).

Regarding co-authorships, if four authors elaborate a particular document and the first two authors are affiliated with University A, the third with University B, and the fourth with a technology start-up C, it produces an interinstitutional network conformed by three nodes mutually connected: two universities (i.e., A and B) and the technology start-up (i.e., C) (Glänzel & De Lange, 1997).

We deepen in the global and individual conceptual and collaboration structures through SNA. The indicators calculated were density and number of communities, for the global network, and betweenness, for each networks' node (i.e., in the co-word analysis, each node is a keyword, in the co-authorship each node is a university).

Density is the proportion of links in a network relative to the total number of links possible. A density of 1 means the network is completely connected. A density closer to 0 reflects a fragmented and disconnected network (Scott, 1988). A node with a higher betweenness, on the other hand, can connect different network clusters and the flow of information between peripherical nodes with a lower degree (Opsahl et al., 2010). Simply put, the higher the betweenness of a node, the higher control that node has over the network, given the flow of information and knowledge flowing through it. In both the co-word and co-authorship analysis, the nodes with the highest betweenness serve as knowledge and relational bridges between clusters. Equation 1 shows how to calculate a network's density.

$$den = 2L/n(n-1) \quad (1)$$

Source: Scott (1988).

Where *L* is the number of links and *n* the number of nodes. Equation 2 shows how to calculate a node's betweenness.

$$C_B(p_k) = \sum_{i<j}^{n} \frac{g_{ij}(p_k)}{g_{ij}}; i \neq j \neq k \quad (2)$$

Source: Opsahl et al. (2010).

Where $g_{ij}$ is the shorter path that links nodes $p_i$ and $g_{ij}(p_k)$ is the shorter path that links nodes $p_i$ and $p_j p_k$.

Analyses were conducted jointly and individually for China and LAC to identify knowledge and institutional bridges at the inter and intraregional levels. The MCA and the SNA followed the approach of Aria and Cuccurullo (2017), Batagelj and Cerinšek (2013), respectively. Both were processed with bibliometrix (2017) for R (R Core Team, 2014). We used Gephi (Bastian et al., 2009) for networks' visualization and indicators calculation. For improving networks' visualization and communities' detection, two algorithms were used: the Circular graph (Six & Tollis, 2006) and the modularity appraisal of Blondel et al. (2008), respectively.

[Table 1]

## 4  Results

More than 3200 keywords were analyzed via MCA and SNA. Figures 1, 2, and 3 present the MCA plots. The MCA of China-LAC explains 38% of the variance by the horizontal and vertical dimensions (Figure 1). Five clusters emerged: i) blue, differentiated by keywords on the management and development of innovation and knowledge in industry; ii) green, composed by keywords related to the development of innovation and technology for sustainable production; iii) purple, related to environmental management and eco-innovation; iv) red, structured by environmental management and green-innovation keywords; and v) orange, related to science, technology and innovation policy. Keywords closer to the origin (0,0) are the less differentiated, such as *technology transfer* or *patents and inventions.*

The MCA of China explains 61% of the variance by both dimensions (Figure 2). Five clusters emerged: i) blue, composed of keywords related to the development of innovation and technology for sustainable production; ii) green, related to science, technology and innovation for industrial development; iii) red, related to IT (Information Technologies) and knowledge management for the supply chain; iv) purple, IT and engineering research; and v) orange, related to technology development and adoption. Keywords such as *product innovation* or *patents and inventions* were the less differentiated.

The MCA of LAC explains 43% of the variance by both dimensions (Figure 3). Four clusters emerged (one cluster was excluded since it was not informative): i) blue, integrated by keywords related to innovation management and public policy and institutions for economic development; ii) red, related to science, technology and innovation for industrial development; iii) green, supply chain management and sustainability; and iv) orange, related to environmental management and green-innovations. The less differentiated keywords were *engineering*, and, again, *patents and inventions.*

[Figure 1]

[Figure 2]

[Figure 3]

Figures 4 and 5 show the overall network indicators. Not even 1% of the China-LAC co-word network is connected (Figure 4). Compared to China's co-word network, LAC's is more connected as its density rose 2% while the former 1%. The number of keyword communities is practically the same: 20 and 21 for China and LAC, respectively.

The China-LAC institutional co-authorship network examined 1,300+ institutions (Figure 5). As discussed above, it shows a reduced density, even less than the co-word network. Here a difference emerged. The institutional co-authorship network in China is denser than that of LAC. That observation is reinforced since the number of communities identified in LAC (294) more than doubles Chinese communities (113).

[Figure 4]

**[Figure 5]**

Figures 6 and 7 present the co-word and institutional co-authorship networks, respectively. Nodes' size is proportional to their betweenness indicator. Tables 2 and 3 present nodes' betweenness indicators of the co-word and institutional co-authorship networks, respectively. The knowledge bridges for both regions are research on *competition* (betweenness: 0.07, e.g., *The influence of Chinese environmental regulation on corporation innovation and competitiveness*), followed by *technological innovation* (0.06, e.g., *Government-subsidized R&D and firm innovation: Evidence from China*), and *commerce* (0.06, e.g., *Service innovation and new product performance*) (Table 2). China shares two out of the three mentioned, with the addition of research on *industry.* (0.06, e.g., *The impact of university-industry collaboration networks on innovation in nanobiopharmaceuticals*). In LAC, there is a considerable difference. The knowledge bridge is the research on *sustainable development* (0.10, e.g., *Eco-innovation in supply chain management*), followed by *commerce* (0.06) also present in the China-LAC and LAC networks, and *environmental impact* (0.06, e.g., *Environmental policy instruments, environmental innovation and the reputation of enterprises*).

Keywords with a higher betweenness in China but absent in LAC, are *patents and inventions* (0.05, e.g., *Managing invention and innovation*), *innovation performance* (0.05), *decision making* (0.03, e.g., *Effects of government financial incentives on firms' innovation performance*), *knowledge management* (0.03, e.g., *Forms of knowledge and modes of innovation*). In contrast, *technological development* (0.05, e.g., *Technological innovation in Brazilian industry: An assessment based on the São Paulo innovation survey*), *research* (0.05, e.g., *Research and innovation in university research groups*), *developing countries* (0.04, e.g., *Developing countries and innovation: Searching for a new analytical approach*), and

*technology transfer* (0.04, e.g., *Effective practices for sourcing innovation*) were highlighted in LAC.

Regarding the institutional co-authorship China-LAC network, most of the top-ten is dominated by Chinese universities, such as that of Tsinghua (0.05), Zhejiang (0.05), and Renmin (0.03) (Table 3). The only two universities from LAC are Sao Paulo (0.05) and Chile (0.02). Besides the aforementioned Chinese institutions, other universities emerged, such as Tongji (0.03), South China (0.05), or Xi'an Jiaotong (0.04). Most of the universities in LAC are located in Brazil, such as Campinas (0.004), and Rio Grande Do Sul (0.004). As opposed to China, universities from Europe also integrated the LAC network (e.g., Kassel in Germany: 0.003 and Murcia in Spain: 0.002).

[Figure 6]

[Figure 7]

[Table 2]

[Table 3]

## 5    Discussion

RIBM boards different spheres from the private sector. While populated clusters were focused on managing and developing STi for industrial purposes, or the implementation of IT in the supply chain, other clusters focused on innovation management in public policy and institutional perspectives for economic development.

Sustainability factors also permeated RIBM in multiples activities related to supply chain management, environmental impact, and the development of green-innovations. China and LAC exhibit different priorities. The most populated cluster of China accentuates the RIBM for the development of STi for sustainable production, while in LAC highlights management and public policy and institutional perspectives for economic development. Two closely related clusters differentiate the RIBM of China over LAC, namely: IT and knowledge management for the supply chain, and IT and engineering research.

Studies conducted in China on the inclusion of technology foresight and road-mapping in national and regional policy for supporting emerging green technologies such as solar cells, support the prominence of the cluster here labeled as *development of innovation and technology for sustainable production* (Huang, Zhang, Guo, Zhu, & Porter, 2014; Li, Zhou, Xue, & Huang, 2015). In those studies, bibliometric insights were vertical inputs in developing regional and national policies.

Previous findings on the comparative advantage of China in *engineering science* supports the differentiation of the cluster labeled as *IT and engineering research* (Harzing & Giroud, 2014). Reasons argued for explaining that advantage consider governmental guidelines for promoting economic development through manufacturing and export orientation, and bountiful resources invested in research and development activities, particularly for electronic technologies (Harzing & Giroud, 2014).

Cetindamar et al. (2009) and Choi et al. (2012) outlined the relevance of *technology policy* for both developed and developing countries, which is also a differentiated cluster for LAC, not so for China. Another point is the incursion of topics formerly prioritized for developed countries such as new product development (Cetindamar et al., 2009), here labeled as *developing science, technology and innovation for industrial purposes,* adding to that the sustainability factors – more noticeable in China than in LAC.

The non-differentiated keyword of *patents and inventions* has to be analyzed carefully. The MCAs denoted *patents and inventions* among the non-differentiated keywords. Despite that, and contrasting with the SNA results, *patents and inventions* was placed among the top-ten keywords in terms of betweenness, which gives it properties of knowledge bridge among related and non-related RIBM keywords clusters, such as *decision making*, *societies and institutions, sales, environmental regulations,* or even *mergers and acquisitions.* For instance, Guan and He (2007) examined a wide range of research-intensive sectors –ranging from biotech and organic chemistry towards semiconductors and optics– when studying citations of STEM articles and published patents in new patents.

We centered the discussion of the co-word SNA in *competition* and *sustainable development* as the keywords with the highest betweenness in each region. *Competition* was not evident in previous RIBM co-word analysis in developing countries (Cortés-Sánchez, 2019). Co-word analysis on RIBM related fields such as strategic management of innovation conceives *competition* into environmental factors along with *market* or *industry* (Keupp et al., 2012)*,* the latter also among the top-ten keywords. In essence, the knowledge bridge of RIBM for both China and LAC is *competition* and its association with environmental factors, and not an internal organizational process.

*Sustainable development* research, on the other hand, has an inverse relation with RIBM and the field of management in general. As mentioned, sustainability factors permeated several clusters both in China and LAC (e.g., *development of innovation and technology for sustainable production* in China, or *supply chain management and sustainability* in LAC). Evidence also pointed to a research shift from early 1990 when sustainability research in business was exclusively focused on economic growth and consumption, to the period 2000-2019 where the three pillars of sustainability were embedded, namely: economic growth, social development, and environmental protection (Jia et al., 2019).

Nevertheless, one of the most comprehensive studies on the research on Sustainable Development Goals (SDGs) placed the cluster labeled as *Green Supply Chains and*

*Management; Manufacturing/Remanufacturing Systems; Cost Analysis and Optimization Models for Waste Management and Recycling* as the 7th cluster and is among the clusters with slow growth during 2015-2018 (Nakamura et al., 2019). The cluster *Maternal, Newborn, and Child Morbidity and Mortality*, conversely, is composed of the most substantial amount of research. However, the method implemented by Nakamura et al. (2019) differs from the one used here. Betweenness appraisal changes the strategic role of a keyword in the co-word analysis completely. Previous RIBM co-word analysis in LAC based on the term's total links instead of betweenness placed *sustainability* in the 5th place of a peripherical cluster (Cortés-Sánchez, 2019).

Compared to other SNA co-authorship studies in the field of management, the institutional co-authorship network here generated and analyzed is denser. Acedo et al. (2006) calculated a density of 0.0002 to a co-authorship network of 10,000+ authors. That indicator comparison deserves attention since Acevedo et al. (2006) analyzed individual authors, not institutions. The low density of both co-word and institutional co-authorship networks reflects an *open structure*. It leads to both advantages and challenges. Open structure networks facilitate and improve the flow of information and knowledge through concepts or institutions linked indirectly (Putman, 2001). That comes with downsides. Communications need to be more refined, sophisticated, and distinguishable than that usually implemented, besides the slow cultivation of relationships based on trust, particularly in institutional collaboration (Prell, 2009). In that line, the institutional co-authorship network of LAC has a lower density and higher fragmentation than China. Unified regional policy for incentivizing local and national collaboration could be an explanatory factor for such differences (Quan Wei et al., 2017). In the case of LAC, each country's higher education system has its research collaborations programs and incentives (Cortés-Sánchez, 2019).

Most Chinese universities are located in mainland China with an exception from Hong Kong (i.e., City University). LAC co-authorship network is integrated mostly by Brazilian institutions (e.g., Campinas, Rio Grande do Sul or Santa Catarina) or state-owned research corporations (i.e., Brazilian Agricultural Research Corporation). It is remarkable the participation of different universities from Chile (i.e., de La Frontera), Germany (i.e., Kassel), Spain (i.e., Murcia), and Colombia (i.e., National University) as collaboration bridges; also, all public. It is predictable since the sample gathered documents from authors affiliated with LAC institutions. Despite that, European public institutions also excel at collaboration bridges in RIBM. Put differently, the collaboration bridges in LAC are, in other words, multi-continental. That supports and exemplifies the fragmented and *open structure* of the LAC network.

RIBM, despite its private spirit, is being generated collaboratively by reputable public institutions, as previously stated by Cortés-Sánchez (2019). Two Chinese and Brazilian institutions figured in the first two places among the top-ten institutions with the highest betweenness indicator: Tsinghua and Sao Paulo, both public institutions. Most of the Chinese institutions in the top-ten (e.g., Fudan, Tsinghua, and Xi'an Jiaotong) are part of the C9 League, a group of nine public institutions which receive 10% of national research investment and produces 20% of the national research output (Times Higher Education, 2011, 2019).

All the institutions from the LAC network previously mentioned are public except for the Mackenzie Presbyterian University, a private university at Sao Paulo. The University of Sao Paulo is one of the leading LAC and Ibero-American institutions. It ranks the 3$^{rd}$ and 118$^{th}$ places in LAC and worldwide in the QS World University Ranking, respectively (Quacquarelli Symonds, 2020). Findings conceive the city of Sao Paulo as the most vibrant place within Brazil and LAC in terms of betweenness, with two institutions among the top-ten analyzed.

In concrete, the underlying structure of the research topics and institutional collaboration on RIBM produced by China and LAC showed that the body of knowledge is going beyond the exclusive studying of the (and for the) private sector. Furthermore, most of the amount and impact of RIBM is being generated by public universities. RIBM for sustainability was the most populated cluster for China and also the keyword with the highest betweenness in LAC. The collective production and sharing of this body of knowledge is crucial for both regions, bringing the negative impact on economic growth and globalization from the standpoint of developed regions due to the COVID-19 pandemic. The ethical and responsible incursion for contributing to the most central SDGs related-research such as *Maternal, Newborn, and Child Morbidity and Mortality,* from the RIBM could be a further line of fruitful societal impact (i.e., consider the cluster (i) in LAC MCA cluster: innovation management and public policy and institutions for economic development). In any case, the consolidation of regional research priorities and collaboration shall consider the earnings and losses of open structure networks. Thus, while the information and knowledge could flow towards institutions linked indirectly, the communication between peers should be more refined, sophisticated, and distinguishable.

## 6    Conclusions

This study investigated the structure of RIBM in China and LAC via co-word and institutional co-authorship. From a broad perspective, the RIBM boards different spheres from the business and management perspectives exclusively, entering into the regional and national STi innovation policy, and institutional perspectives for economic development. The latter is of crucial relevance for LAC. The global development agenda, crystallized in the SGDs, also has imprinted an influence of sustainability. Examples are related to activities of supply chain management, environmental impact, and the development of green-innovations. China's differences lie in clusters permeated by sustainability factors and those of IT and knowledge management for supply chain management.

Non-differentiated keywords such as *patents and inventions* shall not be overlooked since their betweenness properties reveal their structural prominence for the joint dialogue among, apparently, non-related topics. Further, environmental factors related to the market or industry (i.e., *competition*) or the global development agenda (i.e., *sustainable development*) also play a leading role as knowledge bridges. The *open structure* of both China and LAC institutional co-authorship networks shed light on the composition differences. The high fragmentation of LAC network also carried the additional participation of universities from Europe, beyond regional institutions. In essence, public institutions are interconnecting and producing most of the advances in RIBM, despite its private-oriented aim.

Findings presented and discussed here will provide a comprehensive landscape on research fronts and collaboration for private, public, or independent institutions. Institutions could visualize the current state of research and relations for allocating resources to incentivize, diversify, or focalize –according to each institution's mission– their efforts in generating or implementing RIBM in developing regions. Also, to select and invest in further inter-institutional alliances, either formal (i.e., institutional) or informal (i.e., individual). Further studies may consider other regions from the Global South (e.g., Africa and other Asian countries), additional technological innovation-related fields (e.g., STEM) and complementary bibliographic databases/search engines (e.g., Google Scholar, Dimensions, and Microsoft Academic).

**Dataset**

The dataset is available at the following permanent link [URL] or QR code. ***The dataset will be added after the reviewers' decision or will be sent if needed.***

**Acknowledgments**

**Table 1 Scopus' search query and bibliometric summary**

| Fields | Field categories | Countries | Source | Title keywords | Timespan |
|---|---|---|---|---|---|
| Business, management and accounting Decision sciences | Accounting; Business and International Management; Business, Management and Accounting; Industrial Relations; Management Information Systems; Management of Technology and Innovation; Marketing; Organizational Behavior and Human Resource Management; Strategy and Management; Tourism, Leisure and Hospitality Management; Information Systems and Management; Management Science and Operations Research; Statistics, Probability and Uncertainty | China; Argentina; Bolivia; Brazil; Chile; Colombia; Costa Rica; Cuba; Dominican Republic; Ecuador; El Salvador; Guatemala; Honduras; Mexico; Nicaragua; Paraguay; Peru; Puerto Rico; Panama; Uruguay; Venezuela | Articles | Innovation; Innovación; Inovação | 1998-2018 |

| Bibliometric summary | |
|---|---|
| Sources (journals) | 390 |
| Articles | 1,728 |
| Av. Citations per document | 20.53 |

| Most relevant sources | | |
|---|---|---|
| Journal title | Articles | % |
| *Journal of Technology Management and Innovation* | 131 | 7.6% |
| *Journal of Cleaner Production* | 73 | 4.2% |
| *Technological Forecasting and Social Change* | 58 | 3.4% |
| *Technology Analysis and Strategic Management* | 51 | 3.0% |
| *International Journal of Technology Management* | 44 | 2.5% |
| *Research Policy* | 39 | 2.3% |
| *Chinese Management Studies* | 36 | 2.1% |
| *Gestao E Producao* | 34 | 2.0% |
| *Journal of Business Research* | 34 | 2.0% |
| *Technovation* | 3 | 0.2% |

| Authors collaboration | |
|---|---|
| Single-authored documents | 162 |
| Documents per author | 0.52 |
| Authors per document | 1.92 |

| Output per country | | |
|---|---|---|
| Country | Articles | % |
| China | 361 | 37.0% |
| Brazil | 230 | 24.0% |
| Colombia | 47 | 4.9% |

| | | | |
|---|---|---|---|
| Mexico | 41 | | 4.0% |
| USA | 34 | | 3.0% |
| Spain | 32 | | 3.3% |
| Chile | 21 | | 2.0% |
| Argentina | 20 | | 2.0% |
| Peru | 11 | | 1.0% |
| Ecuador | 8 | | 0.8% |

| Keywords | |
|---|---|
| Total | 3,205 |
| Top ten keywords | Articles keyword |
| *Innovation* | 423 |
| *China* | 156 |
| *Competition* | 63 |
| *Commerce* | 61 |
| *Industry* | 56 |
| *Sustainable Development* | 50 |
| *Technological Innovation* | 49 |
| *Technological Development* | 47 |
| *Knowledge Management* | 43 |
| *Patents And Inventions* | 41 |

**Source: elaborated by the authors based Scopus (2019). and processed with bibliometrix (2017)**

**Table 2 Top-10 keywords with the highest betweenness in China-LAC, China, and LAC**

| China-LAC | | China | | LAC | |
|---|---|---|---|---|---|
| Keyword | Betweenness | Keyword | Betweenness | Keyword | Betweenness |
| Competition | 0.072511 | Competition | 0.082973 | Sustainable Development | 0.107377 |
| Technological Innovation | 0.061771 | Commerce | 0.062892 | Commerce | 0.067764 |
| Commerce | 0.061282 | Industry | 0.062624 | Environmental Impact | 0.061371 |
| Industry | 0.047789 | Technological Innovation | 0.058671 | Competition | 0.061144 |
| Patents and Inventions | 0.043955 | Patents and Inventions | 0.053803 | Technological Innovation | 0.059231 |
| Sustainable Development | 0.042204 | Innovation Performance | 0.053674 | Technological Development | 0.052768 |
| Knowledge Management | 0.038221 | Decision Making | 0.038607 | Research | 0.050037 |
| Decision Making | 0.036136 | Knowledge Management | 0.036728 | Developing Countries | 0.048373 |
| Developing Countries | 0.035448 | Economics | 0.032118 | Technology Transfer | 0.043151 |

Source: elaborated by the authors based Scopus (2019) and processed with Gephi (2009)

**Table 3 Top-10 universities with the highest betweenness in China-LAC, China, and LAC**

| China-LAC | | China | | LAC | |
| --- | --- | --- | --- | --- | --- |
| University | Betweenness | University | Betweenness | University | Betweenness |
| Tsinghua University | 0.05581 | Zhejiang University | 0.123097 | University Of Sao Paulo | 0.023467 |
| University Of Sao Paulo | 0.055446 | Tsinghua University | 0.120381 | Universidad De La Frontera | 0.006542 |
| Zhejiang University | 0.052598 | Renmin University Of China | 0.087945 | University Of Campinas | 0.004925 |
| Renmin University Of China | 0.032499 | Tongji University | 0.073659 | Federal University Of Rio Grande Do Sul | 0.004707 |
| Tongji University | 0.032264 | South China University Of Technology | 0.058801 | Universidade Presbiteriana Mackenzie | 0.004487 |
| South China University Of Technology | 0.026491 | Xi'an Jiaotong University | 0.049107 | Brazilian Agricultural Research Corporation (Embrapa) | 0.003776 |
| University Of Nottingham | 0.022868 | City University Of Hong Kong | 0.046412 | Universidade Federal De Santa Catarina | 0.003183 |
| Xi'an Jiaotong University | 0.022322 | Southwestern University Of Finance And Economics | 0.045715 | University Of Kassel | 0.003098 |
| Fudan University | 0.020977 | Fudan University | 0.040461 | University Of Murcia | 0.00278 |
| University Of Chile | 0.020647 | Shanghai Jiao Tong University | 0.038584 | National University Of Colombia | 0.002438 |

Source: elaborated by the authors based Scopus (2019) and processed with Gephi (2009)

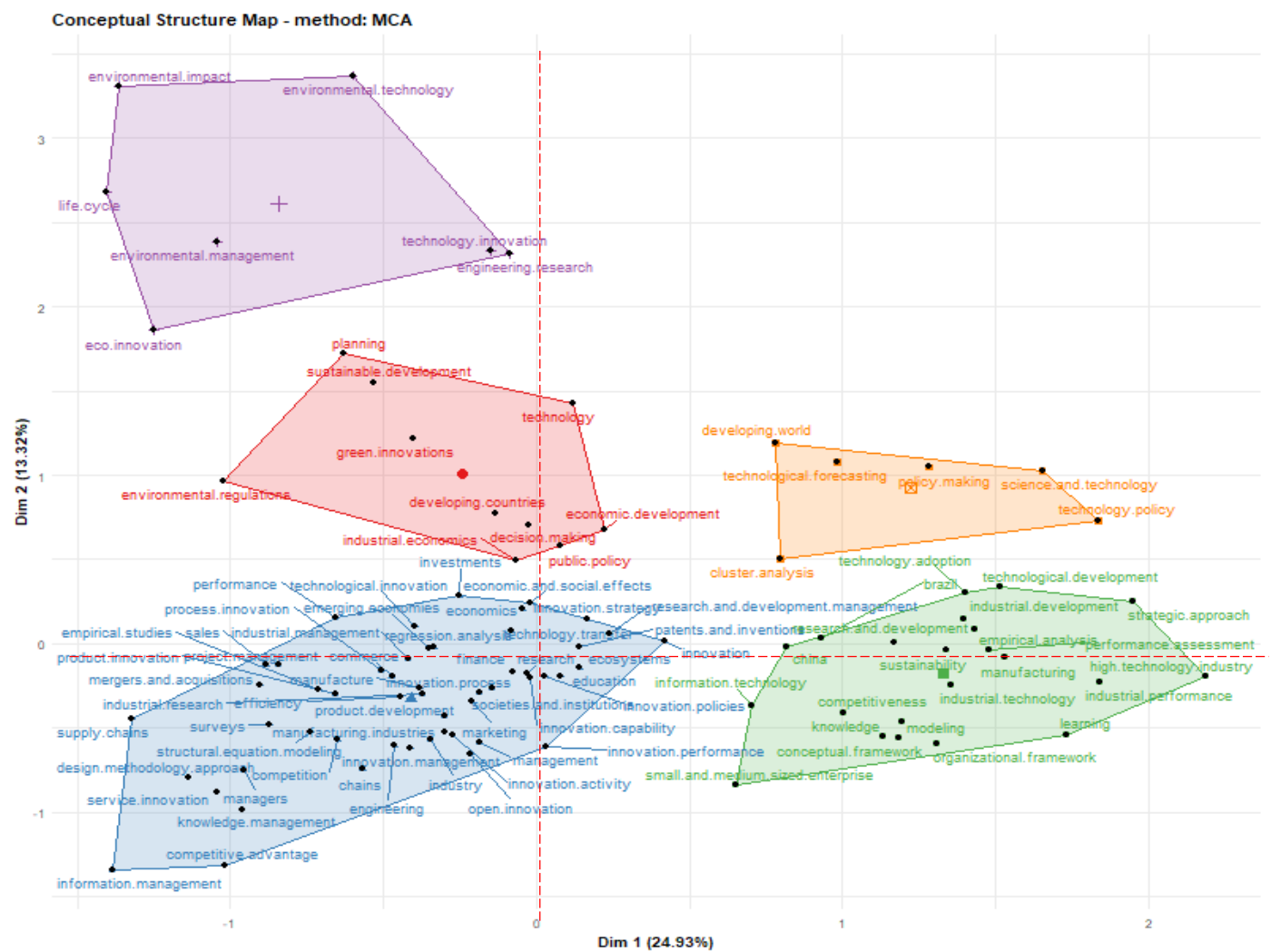

**Figure 1 MCA – China-LAC. Source: elaborated by the authors based Scopus (2019) and processed with bibliometrix (2017)**

**Figure 2 MCA - China. Source: elaborated by the authors based Scopus (2019) and processed with bibliometrix (2017)**

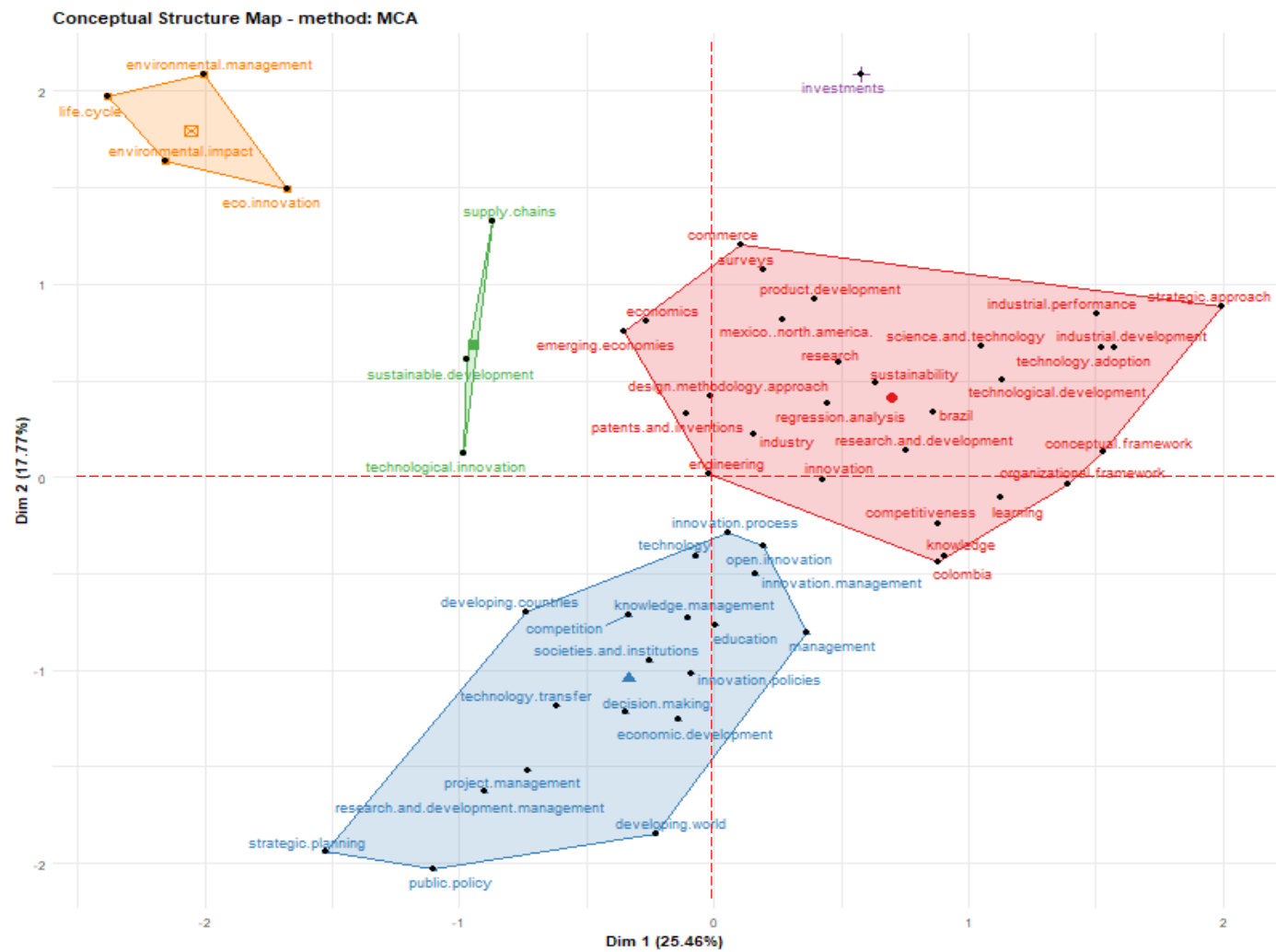

**Figure 3 MCA - LAC. Source: elaborated by the authors based Scopus (2019) and processed with bibliometrix (2017)**

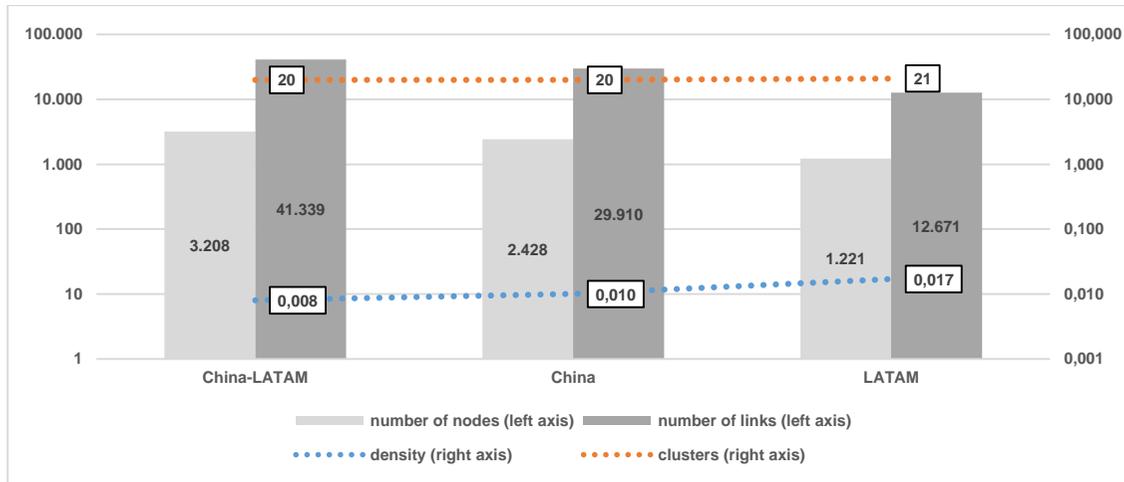

**Figure 4 SNA Network's density, clusters, nodes and links – Co-words. Source: elaborated by the authors based Scopus (2019) and processed with Gephi (2009)**

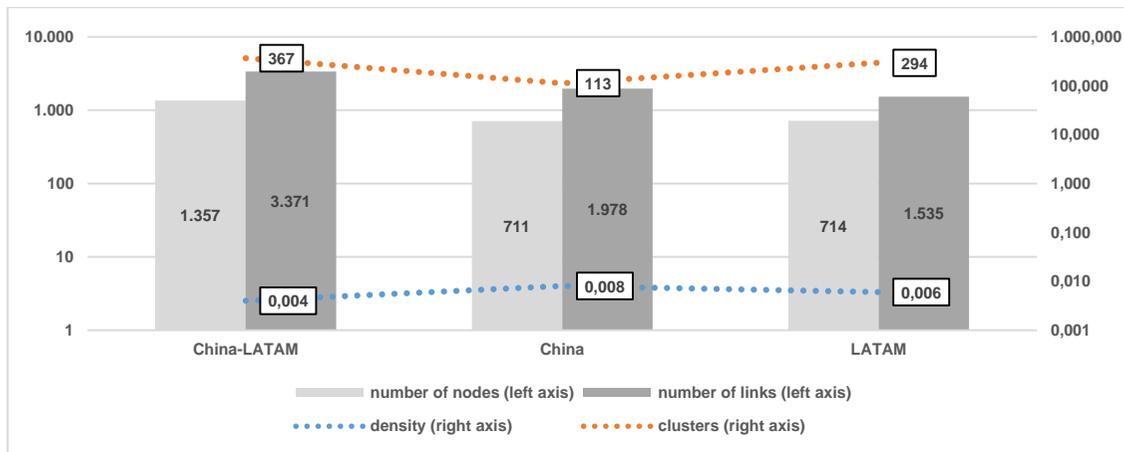

**Figure 5 SNA Network's density, clusters, nodes and links – Institutional co-authorship. Source: elaborated by the authors based Scopus (2019) and processed with Gephi (2009)**

|     a)     |     b)     |     c)     |

**Figure 6 Co-word SNA - China and LAC a); China b); and LAC c). Source: elaborated by the authors based Scopus (2019) and processed with Gephi (2009)**

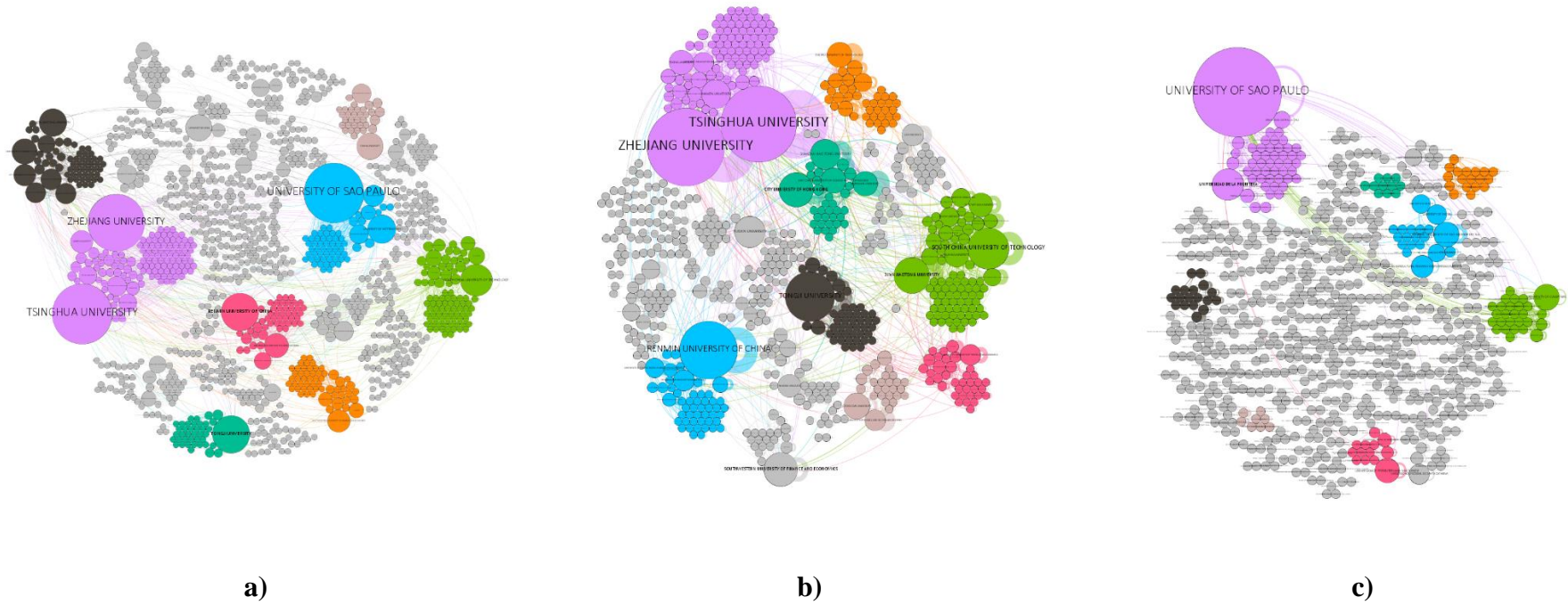

**Figure 7** Institutional collaboration SNA - China and LAC a); China b); and LAC c). Source: elaborated by the authors based Scopus